\documentclass[prb,twocolumn,showpacs,amssymb]{revtex4}
\usepackage{amsmath,graphics,epsfig}
\usepackage{bm}

\newcommand{\mac}{\mathcal}
\newcommand{\tx}{\text}
\newcommand{\ti}{\textit}
\newcommand{\tb}{\textbf}

\newcommand{\dg}{\dagger}

\newcommand{\nn}{\nonumber}
\newcommand{\pat}{\partial}
\newcommand{\pr}{\prime}

\newcommand{\alp}{\alpha}

\newcommand{\Dlt}{\Delta}

\newcommand{\eps}{\epsilon}
\newcommand{\gm}{\gamma}

\newcommand{\tha}{\theta}

\newcommand{\vps}{\varepsilon}
\newcommand{\og}{\omega}

\newcommand{\sig}{\sigma}

\newcommand{\etal}{{\em et al.~}}
\newcommand{\ie}{{i.e.,~}}

\begin{document}
\title{Rashba diamond in an Aharonov-Casher ring}
\author{Xuhui Wang}
\email{xuhui.wang@kaust.edu.sa}
\author{Aurelien Manchon}
\affiliation{Physical Science \& Engineering Division, KAUST, Thuwal 23955-6900,
Kingdom of Saudi Arabia}
\date{\today}

\begin{abstract}
Spin interference due to Rashba spin-orbit interaction (SOI) in a ballistic
two-dimensional electron gas (2DEG) ring conductor submitted to a bias voltage is investigated theoretically. 
We calculate the scattering matrices and differential conductance
with lead-ring junction coupling as an adjustable parameter. Due to the interference of electronic waves
traversing the ring, the differential conductance modulated by both bias voltage 
and SOI exhibits a diamond-shaped pattern, thus termed as \ti{Rashba diamond}.
This feature offers a supplementary degree of freedom to manipulate phase interference.
\end{abstract}
\maketitle

Interference phenomena, such as the celebrated Aharonov-Casher (AC) effect,\cite{ac-effect}
on a low-dimensional ring-shaped conductor patterned on a 2DEG 
with a Rashba\cite{rashba-soi} SOI have  
attracted much attention.\cite{rashba-ring-theory, aronov-lyandageller-1993, rashba-ring-exp} 
Rashba SOI, due to structure inversion symmetry breaking, 
is dominating in quantum wells made of narrow gap semiconductors\cite{nitta-1997-prl-and-engels-1997-prb}
and is among the most popular candidates to phase-coherent spintronic devices.\cite{datta-das,nitta-1999-apl}
Recently, a large (at the order of $10^{-11}~\tx{eV}~\tx{m}$) while tunable Rashba parameter $\alp$
(controled by a gate voltage) 
has been achieved in InGaAs-based 2DEG systems. 
\cite{nitta-1997-prl-and-engels-1997-prb,large-soi-parameter}

We consider, in this letter, a one-dimensional (1D) ring conductor 
fabricated on a 2DEG with a Rashba SOI.
In such a system, electrons experience an effective magnetic field 
$\bm{B}_{eff}\propto\alp\bm{p}\times\hat{\bm{z}}$ that is perpendicular to 
the momentum $\bm{p}$ while in the 2DEG plane. 
Electron waves that traverse the ring along clockwise and 
counterclockwise directions accumulate different phases that depend on  
$\alp$ and the incident energy, which is reflected in the interference patterns of 
the conductance. Most studies were focusing on the 
conductance as a function of gate electric fields (therefore $\alp$) and magnetic fields,
see, Refs.\onlinecite{rashba-ring-exp} and references therein.
Nitta \etal proposed a gate-controlled spin-interference device on a ring 
conductor with a Rashba SOI, \cite{nitta-1999-apl}   
while in this letter, we tune the interference patterns by 
applying a bias (therefore modifying the energy of incident electrons),
thus offering a supplementary degree of freedom to control.
We also address the impact of the lead-ring junction transparency on the interference
\cite{1984-buttiker-etal-pra} which is usually ignored in the AC ring literature. 

\begin{figure}[tbh]
\includegraphics[trim = 6mm 0mm 4mm 0mm, clip, scale=0.4]{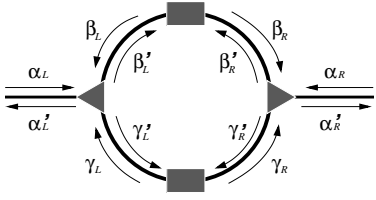}
\caption{A one-dimensional ring connected to two leads. 
The amplitudes and the transport directions of electron waves are 
denoted using arrows. The ring and the leads are made of same 2DEG and the entire 
structure is under an electric field which tunes the Rashba SOI parameter $\alp$.}
\label{fig:ring}
\end{figure}

In Fig. \ref{fig:ring}, the desired ring conductor is connected to two leads. 
At low temperature, when the conducting channel length is comparable to  
the mean free path of electrons, a phase-coherent transport is justified.\cite{datta-book} 
Meanwhile, the Dyakonov-Perel spin relaxation mechanism\cite{dyakonov-perel-pla} 
is reduced in a two-dimensional strip.\cite{malshukov-chao-prb-and-kiselev-kim-prb,nitta-jap-2009}
Since the width of the ring branches is much shorter
than the dimension along transport direction, 
the energy level splitting due to transverse confinement is much
larger than the energy spacing along transport direction, which supports a single-channel 
transport.

A static electric field is applied perpendicularly to the 
2DEG plane and the magnetic field is absent. 
The 1D single-particle Hamiltonian for 
electrons in the ring is\cite{2002-meijer-etal-prb}
(in cylindrical coordinates with an in-plane angle $\phi$)
\begin{align}
H_{1D}
=&-\frac{\hbar^{2}}{2m_{e} a^{2}}\frac{\pat^{2}}{\pat\phi^{2}}
+i\frac{\alp}{a}\left(\frac{1}{2}\hat{\sig}_{\phi}+\hat{\sig}_{r}\frac{\pat}{\pat\phi}\right),
\label{eq:hami-1d-cylind}
\end{align}
where $m_{e}$ is the effective mass of the electrons and $a$ is the radius 
of the ring. In cylindrical coordinates, the Pauli matrices 
are $\hat{\sig}_{\phi}=\hat{\sig}_{y}\cos\phi-\hat{\sig}_{x}\sin\phi$ 
and $\hat{\sig}_{r}=\hat{\sig}_{x}\cos\phi+\hat{\sig}_{y}\sin\phi$. 
The eigenvalues of $H_{1D}$ are given by 
\begin{align}
\vps_{s}=\frac{\hbar^{2}}{2 m_{e} a^{2}}
\left(n-\frac{\Phi_{s}^{AC}}{2\pi}\right)^{2},
\end{align}
where the polarization index $s=\Uparrow(+1)/\Downarrow(-1)$ and the so-called AC phase
\cite{ac-effect} is $\Phi_{s}^{AC}=-\pi+s\pi\sqrt{\og^{2}+1}$, given $\og=2m_{e} a \alp/\hbar^{2}$.
The corresponding normalized wave functions 
are $\Psi_{\Uparrow(\Downarrow)}=\chi_{\Uparrow(\Downarrow)}e^{in\phi}$,
where the spinors are
\begin{align}
\chi_{\Uparrow} = \frac{1}{\sqrt{2}}\left(
\begin{array}{c}
\cos\frac{\tha}{2}\\
\sin\frac{\tha}{2}e^{i\phi}
\end{array}
\right) ,
\chi_{\Downarrow} = \frac{1}{\sqrt{2}}\left(
\begin{array}{c}
\sin\frac{\tha}{2}\\
-\cos\frac{\tha}{2}e^{i\phi}
\end{array}
\right),
\label{eq:eigenvectors}
\end{align}
given $\sin\tha=\og/\sqrt{\og^{2}+1}$.
When an electron of energy $E=\hbar^{2}k^{2}/2m_{e}$ along the transport
direction enters the ring (say, from the left lead), energy conservation requires
\begin{align}
\frac{\hbar^{2}k^{2}}{2m_{e}}=\frac{\hbar^{2}}{2 m_{e} a^{2}}
\left(n-\frac{\Phi_{s}^{AC}}{2\pi}\right)^{2},
\label{eq:eigen-values}
\end{align}
leading to  $n_{s,1}=ak+\Phi_{s}^{AC}/(2\pi)$ and $n_{s,2}=-ak+\Phi_{s}^{AC}/(2\pi)$
for each polarization ($s$). The phase $in_{s,1(2)}\phi$ is the sum of the dynamical 
and the AC phases accumulated when traveling clockwisely (counterclockwisely).

\ti{Scattering matrix.} The transport property of 
a coherent conductor is described by a scattering matrix.\cite{blanter-buttiker-pr} 
We derive a total scattering matrix $\mac{S}$ that 
converts the incoming amplitudes ($\alp_{L}$ and
$\alp_{R}$, see Fig. \ref{fig:ring}) to the outgoing ones ($\alp_{L}^{\pr}$ and
$\alp_{R}^{\pr}$) for each eigenstate in Eq.(\ref{eq:eigenvectors}).
In the ballistic ring without spin-flip scatterings,
we may treat $\Psi_{\Uparrow}$ and $\Psi_{\Downarrow}$
separately,\cite{aronov-lyandageller-1993} and the total conductance is the sum of the 
contributions from these two states. 

On each branch, a scattering matrix is assigned to each spin polarization, \ie
$\mac{S}_{u}$ on the upper branch and $\mac{S}_{d}$ on the lower one
(polarization index omitted for brevity).
Since the spin-flip mechanism is suppressed, traversing 
a ring branch is equivalent to accumulating a phase shift. 
As an example, matching of the wave functions for spinor $\chi_{\Uparrow}$ 
in the upper branch at two junctions gives the 
scattering matrix $\mac{S}_{u}$ satisfying \cite{aronov-lyandageller-1993}
\begin{align}
\left(\begin{array}{c} \beta_{L}\\ \beta_{R}
\end{array}\right)
=\left(\begin{array}{cc}
0 & e^{i n_{2}\pi}\\
e^{-i n_{1}\pi} & 0
\end{array}
\right) \left(\begin{array}{c} \beta_{L}^{\pr}\\
\beta_{R}^{\pr}
\end{array}\right),
\end{align}
and $\mac{S}_{d}=\mac{S}_{u}^{T}$. Two components of the spinor $\chi_{\Uparrow}$
differ by a phase factor that is unimportant,
\cite{aronov-lyandageller-1993} thus neglected. 
Same consideration applies to spinor $\chi_{\Downarrow}$. 
At each junction (triangles as in Fig. \ref{fig:ring}) and for each polarization $s$,
three incoming waves ($\alp_{L(R)}$, $\beta_{L(R)}$ and $\gm_{L(R)}$) 
are scattered to the outgoing ones ($\alp_{L(R)}^{\pr}$, $\beta_{L(R)}^{\pr}$ and $\gm_{L(R)}^{\pr}$)
by a symmetric scattering matrix \cite{1984-buttiker-etal-pra}
\begin{align}
\mac{S}_{L(R)}=\left(%
\begin{array}{ccc}
-\cos\eta & \frac{1}{\sqrt{2}}\sin\eta & \frac{1}{\sqrt{2}}\sin\eta\\
\frac{1}{\sqrt{2}}\sin\eta & -\sin^{2}(\eta/2) & \cos^{2}(\eta/2)\\
\frac{1}{\sqrt{2}}\sin\eta & \cos^{2}(\eta/2) & -\sin^{2}(\eta/2)
\end{array}
\right),
\end{align}
where the junction transparency $\eta\in [0,\pi/2]$: \cite{nazarov-book} 
When $\eta= 0$, the ring is decoupled from the leads; 
when $\eta=\pi/2$, an incoming wave is fully transmitted to two branches with equal probability. 
To keep the essential physics, we consider two identical junctions.

\ti{Transmission probability.} Time-reversal symmetry leads to a
symmetric $\mac{S}$ that satisfies $\mac{S}^{\dg}\mac{S}=1$ as 
required by current conservation. Total conductance 
$G(E,\alp)=(e^{2}/h)\mac{T}(E,\alp)$,  proportional to the total transmission probability 
$\mac{T}=\mac{T}_{\Uparrow}+\mac{T}_{\Downarrow}$, 
accounts contribution from two polarizations. 
When $\eta=\pi/2$, the total transmission probability is
\begin{align}
\mac{T}_{\pi/2} 
= \frac{16(1-\cos\Dlt)\sin^{2}(a k \pi)}
{(1+\cos\Dlt)^{2}+8(1-\cos\Dlt)\sin^{2}(a k \pi)},
\end{align}
where $\Dlt=\pi\sqrt{\og^{2}+1}$ is half of the AC phase difference between two polarizations. 
When $\alp=0$ (SOI vanishes), $\Dlt=\pi$ and  
$\mac{T}_{\pi/2}=2$ can be understood as $\mac{T}_{\pi/2}=1_{\Uparrow}+1_{\Downarrow}$: 
In a fully transparent ring, the transmission probability of each polarization is unity. 
At a weak coupling $\eta=\pi/6$, the total transmission probability is 
\begin{align}
\mac{T}_{\pi/6}
= \frac{16 \xi^{2} (1-\cos\Dlt)\sin^{2}(a k \pi)}
{(\xi^{2}+14\xi\cos(2 a k \pi)-\cos\Dlt)^{2}+4\xi^{2}\sin^{2}(2 a k \pi)}
\end{align}
where $\xi=4\sqrt{3}-7$. It is interesting to notice that the case in 
Ref.\onlinecite{molnar-etal-prb} corresponds to a transparency 
$\pi/3<\eta< \pi/2$.

\ti{I-V curves.} Applying a bias voltage ($V$) across the ring, the current (as a function of 
$V$, temperature ($T$), and $\alp$) is given by an energy integration
\begin{align}
I(V,\alp, T)=\frac{e}{2\pi\hbar}\int_{0}^{+\infty}\mac{T}(E,\alp)
\left[f_{L}(E)-f_{R}(E)\right]dE,\nn
\end{align}
where $f_{L}(E)=f( E-(E_{F}+eV))$ and $f_{R}(E)=f(E-E_{F})$ are the Fermi-Dirac 
distribution functions of the leads and $E_{F}$ is the Fermi energy. 

\begin{figure}[tbh]
\centering
\includegraphics[trim = 0mm 0mm 0mm 0mm, clip, scale=0.7]{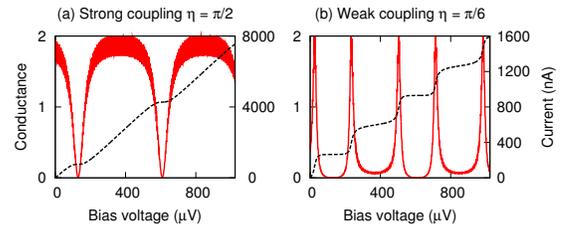}
\caption{(Color online) Differential conductance (left $y$-axis, red solid lines) and I-V curves 
(right $y$-axis, black dashed lines) at  $\eta=\pi/2$ (strong coupling) and at 
$\eta=\pi/6$ (weak coupling).  The conductance is in the unit of $e^{2}/h$. The radius of the ring $a=1~\mu\tx{m}$.
The Rashba parameter $\alp=1.8~\tx{peV}~\tx{m}$,
$E_{F}=75~\tx{meV}$, temperature $T=15~\tx{mK}$,
and the effective mass $m_{e}\approx 0.05 m$.}
\label{fig:dciv}
\end{figure}

Figs. \ref{fig:dciv} show the step-like I-V curves upon scanning the bias voltage,
which is similar to the Coulomb blockade in quantum dot systems, except that the 
plateaus in this AC ring is determined by the destructive interference of 
phases rather than the Coulomb repulsion.  
Fixing $\alp$, decreasing $\eta$ makes 
the steps more developed and the conductance peaks sharper. 
The width of the conductance peak, as broadened by the lead-ring coupling,\cite{aronov-lyandageller-1993,1984-buttiker-etal-pra}   
reflects the lifetime of an electronic state inside the ring: The narrower is the
peak the longer is the lifetime. \cite{1984-buttiker-etal-pra} 
When $\eta$ is small, the total transmission probability 
\begin{align}
\mac{T}_{\eps}\approx 4\eps^{2}\frac{(1-\cos\Dlt)\sin^{2}(a k \pi)} 
{(\cos\Dlt+\cos(2 a k\pi))^{2}},
\label{eq:weak-limit}
\end{align}
where $\eps\approx\eta^{2}/2$ as introduced in Ref. \onlinecite{1984-buttiker-etal-pra}. 
The singularities of the transmission probability in Eq.(\ref{eq:weak-limit}) are determined by 
the solutions of $\cos\Dlt+\cos(2 a k \pi)=0$  
which is exactly the eigenvalue equation Eq. (\ref{eq:eigen-values}). 

\begin{figure}[tbh]
\begin{center}
\includegraphics[trim = 0mm 0mm 0mm 0mm, clip, scale=0.45]{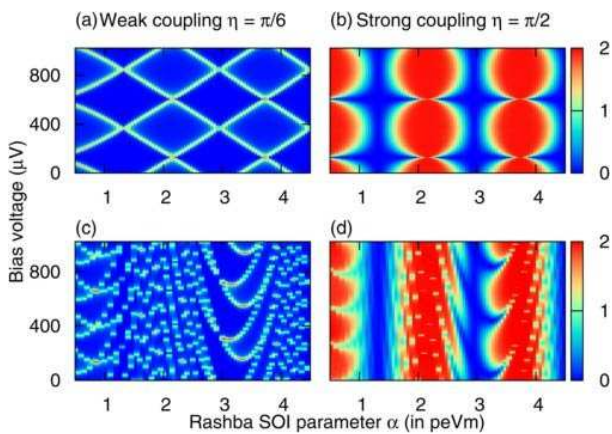}
\caption{(Color online) Differential conductance as a function of $\alp$ (horizontal axis) and 
bias voltage $V$ (vertical axis). The differential conductance is in 
the unit of $e^2/h$. The ring radius is $a=1~\mu\tx{m}$,
temperature $T=15~\tx{mK}$, and $m_{e}\approx 0.05 m$.
In panels (a) and (b), $E_{F}=75~\tx{meV}$. 
In (c) and (d), the carrier density dependence of $\alp$  is taken from Ref.\onlinecite{nitta-jap-2009}.}
\label{fig:rashbadiamond}
\end{center}
\end{figure}

\ti{Rashba diamond.} Fig. \ref{fig:rashbadiamond} shows the differential conductance 
($dI/dV$) modulated by both $\alp$ (horizontal axis) and $V$ (vertical axis).  
The presence of \ti{Rashba diamonds} in panels (a) and (b), reminds us again the 
\ti{Coulomb diamonds} in a system consisting of, e.g. quantum dots. \cite{nazarov-book} 
For a given $\alp$ and $\eta$, the bias fully controles the conductance, thus leading 
to a new degree of freedom to manipulate the interference-induced current modulation.  
The panels (a) and (b) in 
Fig. \ref{fig:rashbadiamond} are for a quantum well based on InGaAs/InAlAs.\cite{nitta-jap-2009}
In the weak coupling regime, the Rashba diamonds are more developed than in the strong coupling limit. 

In a realistic 2DEG, the gate voltage modulates both $\alp$ and carrier density (therefore 
the Fermi level). \cite{ando-rmp} 
We take the experimental data on an InGaAs/InAlAs based 2DEG from Ref.\onlinecite{nitta-jap-2009}
(the 10nm quantum wells in Fig.3 of Ref.\onlinecite{nitta-jap-2009}), from which   
a linear relationship between $\alp$ and Fermi level is established to
calculate the differential conductance.
In panels (c) and (d) of Fig.\ref{fig:rashbadiamond}, \ti{distorted} Rashba diamonds appear as the 
Fermi level is adjusted simultaneously when changing $\alp$. The 
periodicity due topological phase interference (through $\alp$) survives,
which is particularly clear in the strong coupling limit.  

To circumvent the experimental difficulty 
of applying a well defined voltage bias, \cite{leturcq-prl} energy barriers formed using 
gate electrodes at the ring-lead junctions are suggested. The advantage of gate electrode is the
flexibility to tune the height of the barrier thus the junction transparencies. Non-magnetic tunnel 
barriers are good candidates as well since we do not expect significant spin-flip scattering at the 
junctions. Another experimentally relevant geometry consists a ring conductors with
two branches interrupted by tunnel barriers or gate electrodes, as in the electric Aharonov-Bohm
experiment.\cite{vanderwiel-prb}

In conclusion, for an AC ring with a Rashba SOI, we have investigated the interference patterns due to 
topological phases. The scattering matrices, as parametrized by junction transparency, were obtained 
analytically. We have demonstrated the possibility to control 
the differential conductance by both a bias voltage and an electric field. A diamond-shaped pattern 
of the conductance is well developed in the weak coupling regime, which provides a 
supplementary degree of freedom to manipulate spins in the Rashba SOI based 
spintronic devices.

\end{document}